\documentclass[conference]{IEEEtran}
\IEEEoverridecommandlockouts
\usepackage{cite}
\usepackage{amsmath,amssymb,amsfonts}
\usepackage{algorithmic}
\usepackage{graphicx}
\usepackage{textcomp}
\usepackage{multirow}
\usepackage{booktabs}
\usepackage{lipsum}
\usepackage{xcolor}
\usepackage{graphicx}
\usepackage{subcaption}
\usepackage{url}
\def\BibTeX{{\rm B\kern-.05em{\sc i\kern-.025em b}\kern-.08em
    T\kern-.1667em\lower.7ex\hbox{E}\kern-.125emX}}
    \makeatletter

\newcommand{\Rmnum}[1]{\expandafter\@slowromancap\romannumeral #1@}
\makeatother
\makeatletter
\newcommand{\linebreakand}{%
  \end{@IEEEauthorhalign}
  \hfill\mbox{}\par
  \mbox{}\hfill\begin{@IEEEauthorhalign}
}
\makeatother

\usepackage{lipsum}
\begin{document}

\title{LV-UNet: A Lightweight and Vanilla Model for Medical Image Segmentation }

\author{
    \IEEEauthorblockN{\small Juntao Jiang}
    \IEEEauthorblockA{\small
        College of Control Science and Engineering\\
        Zhejiang University\\
        Hangzhou, China\\
        juntaojiang@zju.edu.cn
    }
    \hspace{0.5cm} 
    \and
    \IEEEauthorblockN{\small Mengmeng Wang}
    \IEEEauthorblockA{\small
        College of Computer Science and Technology\\
        Zhejiang University of Technology\\
        Hangzhou, China\\
        mengmewang@gmail.com
    }
    \hspace{0.5cm} 
    \and
    \IEEEauthorblockN{\small Huizhong Tian}
    \IEEEauthorblockA{\small
        The First Clinical Medical College\\
        Guangdong Medical University\\
        Zhanjiang, China\\
        tianhuizhong@gdmu.edu.cn
    }
    \hspace{0.5cm} 
    \and
    \IEEEauthorblockN{\small Lingbo Cheng}
    \IEEEauthorblockA{\small
        College of Control Science and Engineering\\
        Zhejiang University\\
        Hangzhou, China\\
        lingbo1@zju.edu.cn
    }
    \hspace{0.5cm} 
    \and
    \IEEEauthorblockN{\small Yong Liu}
    \IEEEauthorblockA{\small
        College of Control Science and Engineering\\
        Zhejiang University\\
        Hangzhou, China\\
        yongliu@iipc.zju.edu.cn
    }
}

\maketitle

\begin{abstract}
While large models have achieved significant progress in computer vision, challenges such as optimization complexity, the intricacy of transformer architectures, computational constraints, and practical application demands highlight the importance of simpler model designs in medical image segmentation. This need is particularly pronounced in mobile medical devices, which require lightweight, deployable models with real-time performance. However, existing lightweight models often suffer from poor robustness across datasets, limiting their widespread adoption. To address these challenges, this paper introduces LV-UNet, a lightweight and vanilla model that leverages pre-trained MobileNetv3-Large backbones and incorporates fusible modules. LV-UNet employs an enhanced deep training strategy and switches to a deployment mode during inference by re-parametrization, significantly reducing parameter count and computational overhead. Experimental results on ISIC 2016, BUSI, CVC-ClinicDB, CVC-ColonDB, and Kvair-SEG datasets demonstrate a better trade-off between performance and the computational load. The code will be released at \url{https://github.com/juntaoJianggavin/LV-UNet}.

\end{abstract}


\section{Introduction}
Medical image segmentation, which aims to accurately delineate anatomical structures or abnormalities, is essential in disease diagnosis and treatment planning. Unlike traditional methods that relied on manual or semiautomatic methods, which were time-consuming, subjective, and prone to inter-observer variability, computer-assisted techniques can be more efficient.  The advent of deep learning ~\cite{lecun2015deep}, utilizing large annotated datasets and well-designed neural networks to infer pixel-level labels, has revolutionized the field.

Recently, medical imaging solutions have been transitioning from laboratory settings to bedside environments. Termed point-of-care imaging, this approach involves performing testing and analysis directly alongside the patient, with the aim of improving patient care. This shift expects medical AI models to be more real-time and lightweight. For example, task shifting for point-of-care ultrasound (POCUS) includes applications in obstetrics, gynecology, emergency medicine, infectious diseases, and cardiac, abdominal, and pulmonary conditions. At the primary care level in resource-limited settings, it has the potential to expand diagnostic imaging capacity and lead to meaningful health outcomes~\cite{abrokwa2022task}. AI systems for point-of-care require considerations for lightweight design, deployability, and real-time performance. For example, Dulmage et al.~\cite{dulmage2021point} developed a point-of-care and real-time AI system to support clinician diagnosis of a wide range of skin diseases. Lightweight and fast segmentation is crucial in these scenarios, as it ensures the AI system can quickly and accurately delineate areas of interest, facilitating efficient operation on resource-constrained devices and providing precise, timely results.

UNet ~\cite{ronneberger2015u} is a classic medical image segmentation model consisting of an encoder and a decoder with skip-connections, showing a significant improvement compared to previous attempts. UNet's variants like Attention-UNet ~\cite{oktay2018attention}, U-Net++ ~\cite{zhou2018unet++}, ResUnet ~\cite{zhang2018road}, Unet 3+~\cite{huang2020unet} and so on have achieved great success. Due to the impressive performance of Transformers in visual tasks in recent years, Transformer-based U-Net variants like Trans-UNet~\cite{chen2021transunet}, Swin-UNet~\cite{cao2023swin}, MT-UNet~\cite{wang2022mixed} and Unetr~\cite{hatamizadeh2022unetr} and so on have gained wider recognition and have shown remarkable effectiveness. However, these models entail significant computational overhead and a large number of parameters, making them difficult to use in point-of-care applications. Thus, more and more effort has been paid to designing lightweight models in medical image segmentation scenarios. Although existing lightweight medical image segmentation models have made significant progress, they often lack robustness across different datasets. How to address these issues by designing a model with reduced computational overhead, fewer parameters, faster inference time, and maintained performance, while also ensuring robustness across various datasets remains a challenge.

The model design of this paper is based on the following ideas: 1) Utilizing a pre-trained lightweight model in the encoder of a segmentation model is crucial, ensuring the model's robustness across different datasets, especially in small datasets; 2) combining pre-trained and expansion modules forms a straightforward approach to designing lightweight medical image segmentation models; 3) introducing fusible modules enables re-parametrization during the inference phase, which further reduces parameter count and computational load. Based on these, we propose a UNet variant called LV-UNet with a lightweight and vanilla model design containing pre-trained weights from MobileNetv3-Large and fusible expansion modules, and also adopt the fusion method in VanillaNet and improve the deep training strategy. Experiments are conducted on the ISIC 2016~\cite{gutman2016skin},  BUSI~\cite{al2020dataset}, CVC-ClinicDB~\cite{bernal2015wm}, CVC-ColonDB~\cite{bernal2012towards}, and Kvair-SEG datasets~\cite{jha2020kvasir}, whose results show that the presented method achieves a better trade-off between performance and the computational load.

\section{Related Works}
\subsection{Lightweight medical image segmentation models} 
Significant efforts have been devoted to designing lightweight models for medical image segmentation, with a primary focus on variants of U-Net. Valanarasu et al.~\cite{valanarasu2022unext} proposed UNeXt, a convolutional MLP-based network for image segmentation, featuring an initial convolutional stage and a latent MLP stage with tokenized MLP blocks that project convolutional features, reducing parameters and computational complexity while improving segmentation performance by shifting input channels to learn local dependencies. Ruan et al.~\cite{ruan2022malunet} proposed a MAL-UNet that integrates several different modules: DGA for global and local feature extraction, IEA for dataset characterization and sample connectivity enhancement, CAB for multi-stage feature fusion generating channel-axis attention maps, and SAB for spatial-axis attention map generation on multi-stage features. Ruan et al.~\cite{ruan2023ege} proposed EGE-UNet, integrating Group multi-axis Hadamard Product Attention (GHPA) and Group Aggregation Bridge (GAB) modules to extract diverse pathological information and fuse multi-scale features effectively, significantly reducing the model size.

\subsection{VanillaNet}
Pursuing minimalism in design, VanillaNet ~\cite{chen2023vanillanet} adopts a simple architecture that prioritizes computational efficiency and information preservation.  By combining minimalistic operations, batch normalization, and activation functions, VanillaNet achieves competitive performance without excessive complexity. The deep training strategy and the series-informed activation function are keys to maintaining non-linearity in shallow networks. 

\subsection{MobileNetv3}
MobileNetv3 ~\cite{howard2019searching} is a state-of-the-art convolutional neural network designed for mobile and edge devices, combining high performance with efficiency. It builds on MobileNetv1~\cite{howard2017mobilenets} and MobileNetv2~\cite{sandler2018mobilenetv2} by integrating advanced techniques such as depth-wise separable convolutions, inverted residuals with linear bottlenecks, and Squeeze-and-Excitation (SE) blocks. In addition, it introduces the H-Swish activation function, which improves computational efficiency and performance. MobileNetv3 uses neural architecture search (NAS) to optimize its structure for different performance and efficiency needs, resulting in two variants: MobileNetv3-Large for higher accuracy and MobileNetv3-Small for resource-constrained environments. 
\section{Motivation}
\subsection{The Role of Pre-training}
Pre-training in large datasets allows models to learn rich and generalized feature representations, which significantly improves their performance in various tasks. These learned features capture important patterns and nuances in the data that may not be easily obtained through training on smaller datasets or from scratch. Lightweight models have fewer parameters, which can restrict their ability to learn and generalize from data effectively. Therefore, even though introducing complex changes to model architectures might seem beneficial, these modifications often do not provide enough improvement to outweigh the substantial performance boost gained from using pre-trained models.  On the other hand, while complex model structures might offer improvements on specific datasets, such benefits are usually inconsistent. Relying on pre-trained weights is a more reliable approach to ensure the model's overall performance and robustness.
\subsection{Efficient Model Deployment through Re-parametrization}
During the training phase, a more complex model can be utilized to ensure optimal learning and feature extraction. Once training is complete, the model undergoes re-parametrization for the inference phase. This involves simplifying the model by compressing and fusing parameters, such as merging multiple convolutional filters into fewer ones. This re-parameterized model maintains performance while significantly reducing computational and memory requirements, ensuring efficient deployment in resource-constrained environments.
\section{Methods}
\subsection{Overall Architecture}
The model architecture of LV-UNet contains an encoder, a decoder, and final output layers, which can be seen in Fig.1. The encoder includes pre-trained blocks from MobileNetv3-Large, as well as fusible encoder modules. The decoder consists of fusible decoder modules. There are skip-connections between the corresponding layers in the encoder and decoder, and to avoid significantly increasing the number of parameters, element-wise addition is used instead of concatenation. The number of channels in each stage is 16, 24, 40, 80, 160, 240 and 480.

\begin{figure*}[htbp]
\centering
\centerline{\includegraphics[width=17cm]{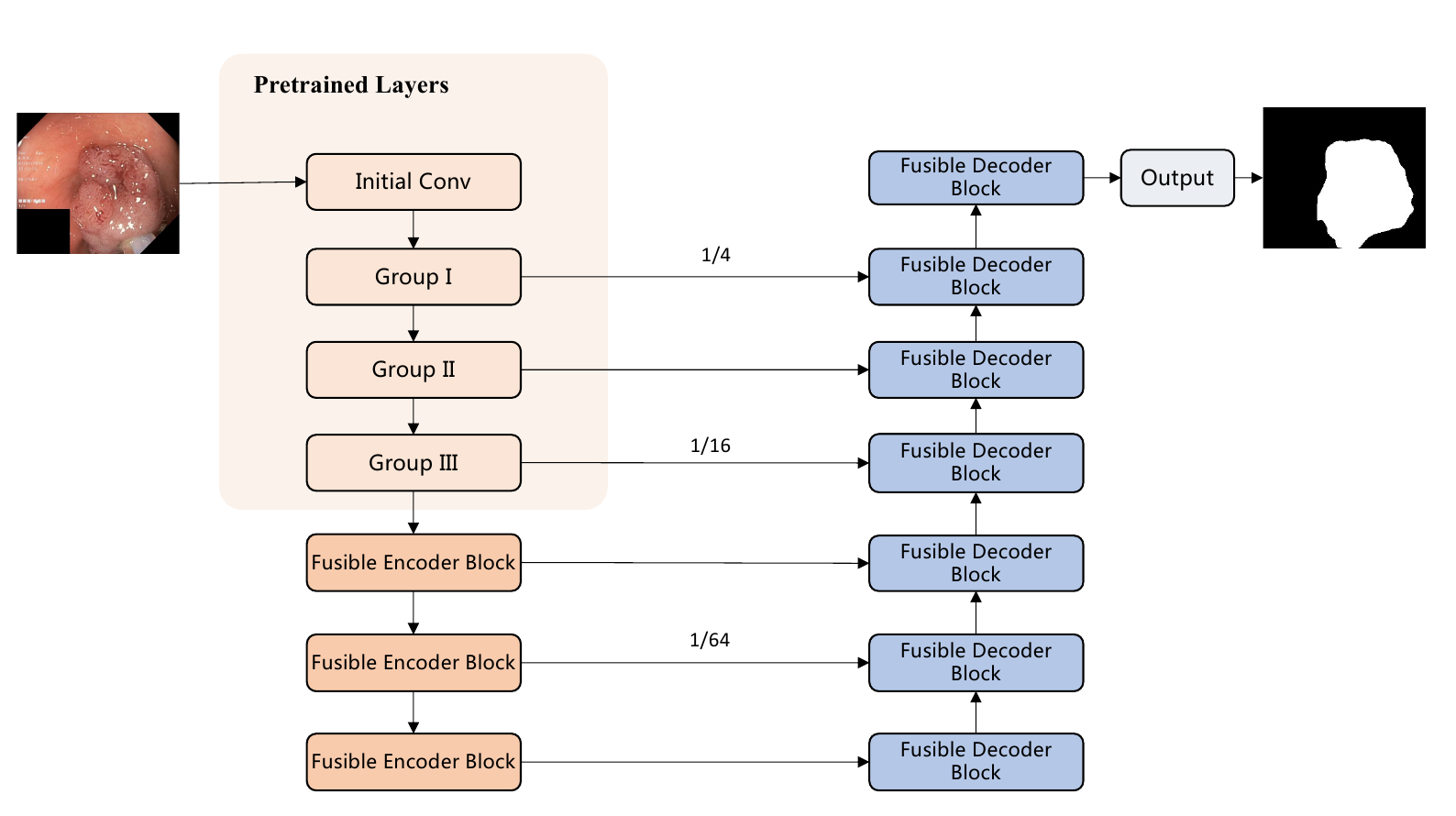}}
\caption{The architecture of LV-UNet: the basic modules include pre-trained MobileNetv3-Large blocks(the initial convolution stage and the group \Rmnum{1} to \Rmnum{3} (the first inverted residual block to ninth), fusible encoder blocks, fusible decoder blocks, skip-connections, and the output block.}
\end{figure*}
\subsection{Pre-trained Modules in Encoder}
LV-UNet uses MobileNetv3-Large from the initial convolutional stage to the ninth inverted residual block for the encoder block of LV-UNet. The initial convolutional stage receives the input image and performs basic feature extraction, and the following layers contain inverted residual blocks with depth-wise separable convolution layers, different convolution kernels and activation functions, progressively extracting more abstract features. We name inverted residual blocks in the MobileNetv3-Large model as $r1$ to $r15$. These inverted residual blocks are grouped as follows: $r1$ and $r2$ as the first group (group \Rmnum{1}) , $r3$, $r4$, and $r5$ as the second group (group \Rmnum{2}), and $r6$, $r7$, $r8$, and $r9$ as the third group (group \Rmnum{3}). We do not use the entire feature extraction part of the MobileNetv3-Large model, to allow flexible design of subsequent expansion modules. 

\subsection{Fusible Modules}
\subsubsection{Architecture} 
The training mode of a fusible encoder block contains a convolutional layer using a $1 \times 1$ kernel with stride 1, a batch normalization operation, a Leaky ReLU function, a convolutional layer using a $1 \times 1$ kernel, a pooling layer and a designed activation layer. The resolutions of output features of an encoder block are half of the input. In the deployment mode, the block only contains a convolutional layer, a pooling layer, and the designed activation layer. A fusible decoder block is similar, where an upsampling operation replaces the pooling operation. The resolutions of output features of a decoder block are two times of the input. The architectures of the fusible blocks in the training and deployment modes are shown in Fig.2. The method to transfer the training mode to the deployment mode during the inference stage will be introduced in the section "Re-parametrization and Deployment Mode".

\begin{figure}[h]
  \centering
  \begin{subfigure}{0.4\textwidth}
    \centering
    \includegraphics[width=\linewidth]{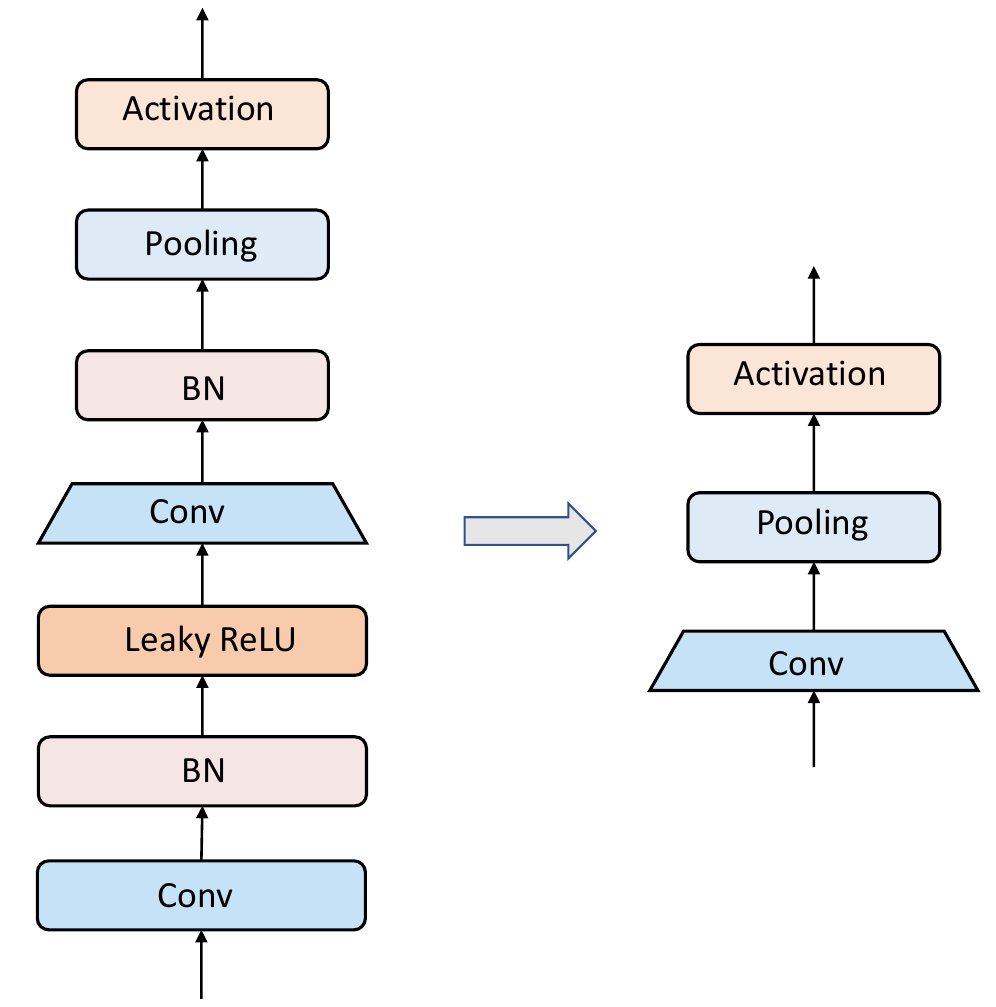}
    \caption{The architecture of the fusible encoder block in the training and deployment modes.}
    \label{fig:sub1}
  \end{subfigure}

  \begin{subfigure}{0.4\textwidth}
    \centering
    \includegraphics[width=\linewidth]{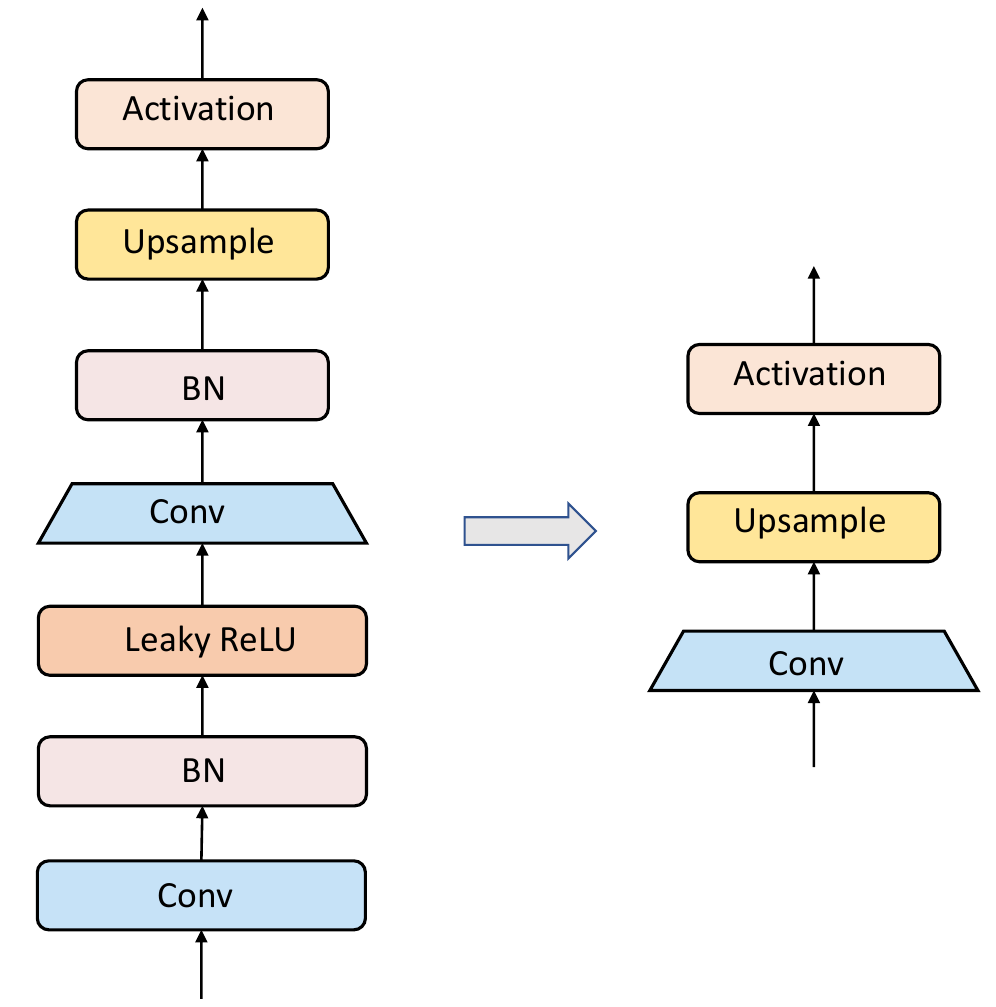}
    \caption{The architecture of the fusible decoder block in the training and deployment modes.}
    \label{fig:sub2}
  \end{subfigure}
  \caption{The architecture of the fusible blocks in the training and deployment modes}
  \label{fig:fig}
\end{figure}
\begin{figure*}[htbp]
\centering
\centerline{\includegraphics[width=18cm]{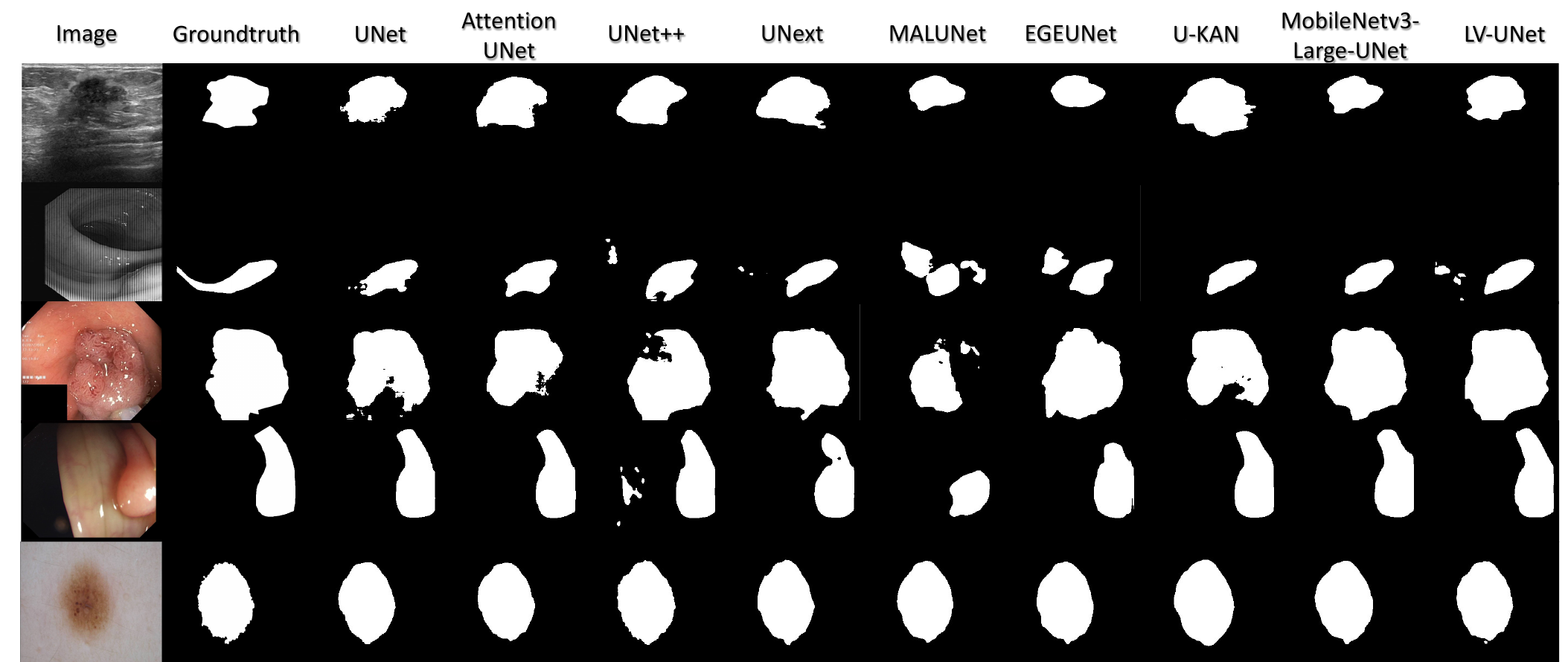}}
\caption{Example visualizations of segmentation results of different models.}
\end{figure*}
\subsubsection{Nonlinear Activation Layer}
The nonlinear activation function in LV-UNet follows the design of series-informed activation functions in VanillaNet. Consider an input feature tensor, denoted as $x \in \mathbb{R}^{H \times W \times C}$, where $H$, $W$, and $C$ represent the height, width, and number of channels, respectively. The activation function is formulated as follows:
\begin{equation}
A_s\left(x_{h, w, c}\right)=\sum_{i, j \in\{-n, n\}} a_{i, j, c} A\left(x_{i+h, j+w, c}+b_c\right)
\end{equation}
where $A_s\left(x_{h, w, c}\right)$ denotes the output of the activation function at position $(h, w, c)$ in the input tensor $h$ ranges from $1$ to $H$, $w$ ranges from $1$ to $W$, and $c$ ranges from $1$ to $C$. The function aggregates inputs from neighboring positions $(i+h, j+w, c)$ within a range defined by $-n$ to $n$. The weights $a_{i, j, c}$ control the contribution of each input, and $A(\cdot)$ represents the activation function applied to each input, typically a nonlinear function like ReLU. Furthermore, $b_c$ represents a bias term specific to each channel $c$. By utilizing this formulation, LV-UNet can capture and incorporate contextual information from neighboring positions, enabling the model to learn and represent complex patterns and nonlinear relationships within the input data.
\subsubsection{Re-parametrization and Deployment Mode}
VanillaNet~\cite{chen2023vanillanet} adopts a deep training strategy in the training process, and it is also applied in the LV-UNet training. At the start of the training process, two convolutional layers utilize an activation function. During the training epochs, the activation function is progressively adjusted to resemble an identity mapping. At the end of the training, the two convolutions can seamlessly combine into a single convolution, resulting in reduced inference time. We use a Leaky ReLU function between two convolutional layers to control it:

\begin{equation}
\text { Leaky } \operatorname{ReLU}(x)=\max (a x, x),\end{equation}

\begin{equation}a=1-cos(\frac{\pi e}{2E})
\end{equation}

where $e$ is the current epoch and $E$ is the number of deep training epochs. At the beginning of the training, the activation function is a ReLU function, while at the end of the training, it is an identity mapping function. The nonlinearity gradually decreases. Unlike the design in ~\cite{chen2023vanillanet} that defines $a$ as $a=\frac{e}{E}$, the designed method can make the transition from linear to nonlinear show a trend of starting slow and accelerating, which aids in the convergence of training.

Following the method used in VanillaNet's training, the weight and bias matrices of a single convolution merged from the batch normalization layer and their preceding convolution in fusible blocks are:
\begin{equation}
W_i^{\prime}=\frac{\gamma_i}{\sigma_i} W_i, B_i^{\prime}=\frac{\left(B_i-\mu_i\right) \gamma_i}{\sigma_i}+\beta_i
\end{equation}
where $W \in R^{C_{\text{out}} \times\left(C_{\text{in}} \times k \times k\right)}$, $B \in R^{\text{out}}$ are the weight and bias matrices of the convolutional kernel, $\gamma, \beta, \mu, \sigma \in R^{\text{out }}$, are the scale, shift, mean and variance in batch normalization, subscript $i \in\left\{1,2, \ldots, C_{out}\right\}$ denotes the value in $i$-th output channels.
 
Then the two 1$\times$1 convolution layers in the fusible blocks can be merged as a single convolution layer:
\begin{equation}
\begin{aligned}
y=W^1 *\left(W^2 * x\right)=W^1 \otimes W^2 \otimes \operatorname{im} 2 \operatorname{col}(x)\\=\left(W^1 \otimes W^2\right) * X
\end{aligned}
\end{equation}

where $*$ means the convolution operation, $\otimes$ represents the matrix multiplication, $W^1$ and $W^2$ are the weight matrix of two convolution layers, and $X \in$ $R^{\left(C_{i n} \times 1 \times 1\right) \times\left(H^{\prime} \times W^{\prime}\right)}$ is derived from the im2col operation to transform the input into a matrix corresponding to the shape of the kernel. 

By using the deep training strategy, LV-UNet can be switched to deployment mode in the inference stage, which reduces parameters and computational requirements. 

\section{Experiments}

\subsection{Datasets}
\begin{table*}[h]
    \caption{Comparison Experimental Results on ISIC 2016, BUSI, CVC-ClinicDB, CVC-ColonDB and Kvasir-SEG datasets}
    \label{comp}
    \centering
    \setlength{\tabcolsep}{3.0mm}
    \begin{tabular}{c|c|c|c|c|c|c|c|c|c|c}
    \hline
 \multirow{2}{*}{\textbf{Methods}}&\multicolumn{2}{|c|}{\textbf{ ISIC 2016 }} & \multicolumn{2}{|c|}{\textbf{ BUSI}} &  \multicolumn{2}{|c|} {\textbf{ CVC-ClinicDB}} &  \multicolumn{2}{|c|} {\textbf{CVC-ColonDB}} &  \multicolumn{2}{|c} {\textbf{ Kvasir-SEG}} \\
   \cline{2-11} &\textbf{IoU} &\textbf{Dice} & \textbf{IoU} &\textbf{Dice}& \textbf{IoU} &\textbf{Dice} &   \textbf{IoU} &\textbf{Dice}&   \textbf{IoU} &\textbf{Dice}\\
        \hline
  
UNet~\cite{ronneberger2015u}& 0.8401& 0.9116&0.6218 & 0.7575 &0.7649 &0.8638&0.7425&0.8387&0.7327&0.8440\\
Attention-UNet~\cite{oktay2018attention}&  0.8430& 0.9137 & 0.6445&0.7755 & 0.7955 &0.8839&0.7523&0.8487& 0.7449&0.8512\\
UNet++~\cite{zhou2018unet++}&0.8375 & 0.9097 &0.6211 &0.7568 &0.7951 & 0.8832& 0.7246&0.8294& 0.7565&0.8598\\
UNeXt~\cite{valanarasu2022unext}& 0.8505& 0.9181&0.6122 &  0.7437&  0.7709 &0.8660& 0.6613&0.7774&0.6880&0.8115\\
MALUNet~\cite{ruan2022malunet}& 0.8374& 0.9098 & 0.5875& 0.7318 &0.4994 & 0.6565&0.5361& 0.6860&0.5724&0.7219\\
EGEUNet~\cite{ruan2023ege}&0.8380 & 0.9103& 0.5571&  0.6949& 0.5190&0.6774& 0.5744&0.7098 &0.5342& 0.6928\\
U-KAN~\cite{li2024u}& 0.8517&0.9192 & 0.6368&  0.7668& 0.8243 &0.9017&0.7367&0.8353&0.7338& 0.8438\\
MobileNetv3-Large-UNet~\cite{howard2019searching} & \textbf{0.8602}& \textbf{0.9240}&0.6226 & 0.7504 &0.8347 &0.9087&0.7802&0.8700&0.7804& 0.8747\\
        \hline
        \textbf{LV-UNet} & 0.8595& 0.9235& \textbf{0.6489}&  \textbf{0.7785}& \textbf{0.8514} & \textbf{0.9183}&0.7964&0.8808&\textbf{0.8069}& \textbf{0.8917}\\
        \hline
          \shortstack{\textbf{LV-UNet} \\ \textbf{(deployment)}} &  0.8593& 0.9234& 0.6347 &  0.7642&0.8272 &0.9045&\textbf{0.7975}&\textbf{0.8811}&0.7956& 0.8840\\
        \hline
    \end{tabular}
\end{table*}

\subsubsection{ISIC2016} ISIC 2016~\cite{gutman2016skin} is a dataset of skin lesions images for Skin Lesion Analysis toward Melanoma Detection Challenge at the International Symposium on Biomedical Imaging (ISBI) 2016. In this paper, we used the lesion segmentation dataset, which contains original images and binary lesion masks. The training set has 900 pairs of original images and corresponding masks. In the testing set, there are 379 pairs of images and masks. The training set is split into a new training set and a validation set with a ratio of 8:2. 

\subsubsection{BUSI} Breast Ultrasound Images Dataset (BUSI)~\cite{al2020dataset} is a dataset of breast ultrasound images collected from women between 25 and 75 years old. The dataset consists of 780 pairs of images in normal, benign, and malignant, with corresponding masks. Unlike the setting in~\cite{valanarasu2022unext}, we include all three classes in the dataset, and the test set is split with a ratio of 0.2. Then the rest of the dataset is split into the training set and the validation set with a ratio of 8:2.

\subsubsection{CVC-ClinicDB} CVC-ClinicDB~\cite{bernal2015wm} is a database of frames extracted from colonoscopy videos that contain several examples of polyps. The dataset consists of 612 pairs of images and corresponding masks. The test set is split with a ratio of 0.2. Then the rest of the dataset is split into the training set and the validation set with a ratio of 8:2.
\subsubsection{CVC-ColonDB}The CVC-ColonDB dataset~\cite{bernal2012towards} includes 380 images, each with a resolution of 574$\times$500 pixels. These images were selected from a pool of 12,000 images obtained from 15 short colonoscopy videos, with only these 380 images being annotated.  The test set is split with a ratio of 0.2. Then the rest of the dataset is split into the training set and the validation set with a ratio of 8:2.

\subsubsection{Kvasir-SEG}
The Kvasir-SEG dataset~\cite{jha2020kvasir} comprises 1000 polyp images and their corresponding ground truth masks with resolutions ranging from 332$\times$487 to 1920$\times$1072 pixels. The test set is split with a ratio of 0.2. Then the rest of the dataset is split into the training set and the validation set with a ratio of 8:2.

\subsection{Implementation Details}

The experiments are all performed on PG500-216(V-100) with 32 GB memory. The total training epochs are 300, and the batch size in training, validation, and testing is 8. The resolution of the input images is resized to 256$\times$256. The optimizer used is ADAM \cite{kingma2014adam}. The initial learning rate is 0.001, and a CosineAnnealingLR \cite{loshchilov2016sgdr} scheduler is used. The minimum learning rate is 0.00001. Only rotation by 90 degrees clockwise for random times, random flipping, and normalization methods are used for data augmentation. The evaluation metric in the validation is $IOU$. The loss function used is a mixed loss that combined binary cross-entropy (BCE) loss and dice loss \cite{milletari2016v}:$$\mathcal{L}=0.5 B C E(\hat{y}, y)+D i c e(\hat{y}, y)$$

We also train, validate and test several models like UNet ~\cite{ronneberger2015u}, Attention UNet ~\cite{oktay2018attention}, UNet++~\cite{zhou2018unet++}, UNeXt~\cite{valanarasu2022unext}, MALUNet~\cite{ruan2022malunet}, EGE-UNet~\cite{ruan2023ege}, U-KAN~\cite{li2024u} and MobileNetv3-UNet (UNet with MobileNetv3 as the backbone) for comparison. UNet uses double convolution layers in each stage. The dimensions of each stage in UNet, Attention UNet and UNet++ are 32, 64, 128, 256 and 512. UNeXt, MLUNet, EGE-UNet, and U-KAN use the original design in their papers. MobileNetv3-UNet adds the output of each encoder stage to the blocks in the decoder.

For other experiments, we select the model that performs the best in the validation set for inference on the test set. However, the Leaky ReLU only transitions to an identity mapping function at the end of the training. Therefore, we choose the model from the last epoch for inference on the test set, for the LV-UNet using the deep training strategy. 

\subsection{Results}
The experimental results of the comparison can be seen in Table \Rmnum {1}, using $IOU$ and $Dice$ as metrics, from which we can see that LV-UNet shows competitive performances compared to baselines. In addition, the deployment mode of LV-UNet can still achieve good results while reducing parameters a lot. In Table \Rmnum {2}, the number of parameters and FLOPs of different models are compared, which shows that the model sizes and computation complexities of LV-UNet and its deployment mode are suitable for edge device and point of care scenarios. Visualizations of some segmentation results can be seen in Fig.3.
\begin{table}[h]
    \caption{Parameters and FLOPs of Different Models }
    \label{comp}
    \centering
    \setlength{\tabcolsep}{4mm}
    \begin{tabular}{c|c|c}
    \hline
\textbf{Methods}&\textbf{ Parameters }  &  \textbf{FLOPs}  \\
        \hline
        UNet~\cite{ronneberger2015u}  & 7.8M & 12.15G  \\
        Attention-UNet~\cite{oktay2018attention}  & 8.7M &16.79G \\
        UNet++~\cite{zhou2018unet++}  &  9.2M & 34.77G \\
        UNeXt~\cite{valanarasu2022unext}  & 1.5M &  0.58G \\
        MALUNet~\cite{ruan2022malunet}  & 0.2M &  0.09G \\
        EGEUNet~\cite{ruan2023ege}  & 53.4K & 0.08G \\
        U-KAN~\cite{li2024u}   & 25.4M & 10.5G \\
        MobileNetv3-Large-UNet~\cite{howard2019searching}  & 2.0M &  0.51G \\
        \hline
        \textbf{LV-UNet} & 0.9M  & 0.22G   \\
        \hline
          \textbf{LV-UNet (deployment)} & 0.5M & 0.20G   \\
        \hline
    \end{tabular}
\end{table}
\begin{table}[h]
    \caption{Experimental Results of Different Combinations for Pre-trained Layers}
    \centering
    \setlength{\tabcolsep}{1.7mm}
    \begin{tabular}{c|c|c|c|c|c|c}
    \hline
\fontsize{7}{7}\multirow{2}{*}{\textbf{Methods}}& \multicolumn{2}{|c|}{\fontsize{7}{7}\textbf{ BUSI}} &  \multicolumn{2}{|c|} {\fontsize{7}{7}\textbf{ CVC-ClinicDB}} &  \multicolumn{2}{|c} {\fontsize{7}{7}\textbf{ Kvasir-SEG}} \\
\cline{2-7}  &\fontsize{7}{7}\textbf{IoU} &\fontsize{7}{7}\fontsize{7}{7}\textbf{Dice} & \fontsize{7}{7}\textbf{IoU} &\fontsize{7}{7}\textbf{Dice}& \fontsize{7}{7}\textbf{IoU} &\fontsize{7}{7}\textbf{Dice} \\
        \hline
  
Combination \Rmnum{1} & 0.6308& 0.7548&\textbf{0.8531} &  \textbf{0.9197} &0.8063 &0.8901\\
Combination \Rmnum{2}&\textbf{0.6489} &\textbf{0.7785} & 0.8514& 0.9183 &\textbf{0.8069} &\textbf{0.8917}\\
Combination \Rmnum{3}&0.6332 &0.7625 &0.8324&  0.9080 &0.7971 &0.8856\\

        \hline
    \end{tabular}
\end{table}
\begin{table}[h]
    \caption{Parameters and FLOPs of Different Combinations for Pre-trained Layers}
    \centering
    \setlength{\tabcolsep}{6mm}
    \begin{tabular}{c|c|c}
    \hline
\textbf{Methods}&\textbf{ Parameters }  &  \textbf{FLOPs}  \\
        \hline
        Combination \Rmnum{1}  & 2.8M & 0.34G  \\
        Combination \Rmnum{2}   & 0.9M &0.22G \\
        Combination \Rmnum{3}  &  0.8M & 0.17G \\
        \hline
    \end{tabular}
\end{table}

\subsection{Ablation Studies}

\textbf{The Number of Pre-trained Layers.} We explore the number of pre-trained layers of MobileNetv3-Large used in LV-UNet, and propose three possible combinations in the encoder: 
\begin{itemize}\item[$\bullet$] Combination \Rmnum{1}: From the initial convolutional stage to the inverted residual block $r14$ in MobileNetv3-Large and two fusible encoder blocks\end{itemize}
\begin{itemize}\item[$\bullet$] Combination \Rmnum{2}: From the initial convolutional stage to  the inverted residual block $r9$ in MobileNetv3-Large and three fusible encoder blocks\end{itemize}
\begin{itemize}\item[$\bullet$] Combination \Rmnum{3}: From the initial convolutional stage to the inverted residual block $r5$ in MobileNetv3-Large and four fusible encoder blocks\end{itemize}
\begin{table}[h]
    \caption{Experimental Results of Different Numbers of  Series Informed Activation Functions}
    \label{comp}
    \centering
    \setlength{\tabcolsep}{2.0mm}
    \begin{tabular}{c|c|c|c|c|c|c}
    \hline
\fontsize{7}{7}\multirow{2}{*}{\textbf{Methods}}& \multicolumn{2}{|c|}{\fontsize{7}{7}\textbf{ BUSI}} &  \multicolumn{2}{|c|} {\fontsize{7}{7}\textbf{ CVC-ClinicDB}} &  \multicolumn{2}{|c} {\fontsize{7}{7}\textbf{ Kvasir-SEG}} \\
 \cline{2-7}  &\fontsize{7}{7}\textbf{IoU} &\fontsize{7}{7}\fontsize{7}{7}\textbf{Dice} & \fontsize{7}{7}\textbf{IoU} &\fontsize{7}{7}\textbf{Dice}& \fontsize{7}{7}\textbf{IoU} &\fontsize{7}{7}\textbf{Dice} \\
        \hline
  
0 & 0.6147&0.7506 &  0.8013& 0.8883 &  0.7891&0.8803\\
1 & \textbf{0.6489}& \textbf{0.7785}&0.8514 & 0.9183 &\textbf{0.8069} &\textbf{0.8917}\\
2&0.6402 &0.7731 & 0.8369& 0.9098 &0.7695 &0.8649\\
3& 0.6370&0.7611 &\textbf{0.8541} & \textbf{0.9198} &0.8015 &0.8847\\
4& 0.6423&0.6423 & 0.8495& 0.9178 & 0.7755&0.8707\\
        \hline
    \end{tabular}
\end{table}
The comparison results can be seen in Table \Rmnum {3}, and the parameters and FLOPs can be seen in Table \Rmnum {4}. The results show that the LV-UNet setting (combination \Rmnum {2}) can achieve better performance on BUSI and Kvasir-SEG datasets and achieve a very close result on CVC-ClinicDB dataset. Meanwhile, compared to combination \Rmnum {1}, combination \Rmnum {2} reduces the model size and computational complexities.

\textbf{The Number of Series-informed Activation Functions.}
In experiments in~\cite{chen2023vanillanet}, the increase of the number of series-informed activation functions can lead to improvements in VanillaNet's classification accuracy. However, in our experiments, no significant relationship was observed between the number of series-informed activation functions and the segmentation accuracy of LV-UNet. However, the results suggest that the use of an appropriate number of series-informed activation functions may potentially enhance segmentation accuracy. The comparison results can be seen in Table \Rmnum {5}.
\begin{table}[h]
    \caption{Comparison of Adding and Concatenation Methods for Skip-Connections}
    \label{comp}
    \centering
    \setlength{\tabcolsep}{2.0mm}
    \begin{tabular}{c|c|c|c|c|c|c}
     \hline
\fontsize{7}{7}\multirow{2}{*}{\textbf{Methods}}& \multicolumn{2}{|c|}{\fontsize{7}{7}\textbf{ BUSI}} &  \multicolumn{2}{|c|} {\fontsize{7}{7}\textbf{ CVC-ClinicDB}} &  \multicolumn{2}{|c} {\fontsize{7}{7}\textbf{ Kvasir-SEG}} \\
\cline{2-7}  &\fontsize{7}{7}\textbf{IoU} &\fontsize{7}{7}\fontsize{7}{7}\textbf{Dice} & \fontsize{7}{7}\textbf{IoU} &\fontsize{7}{7}\textbf{Dice}& \fontsize{7}{7}\textbf{IoU} &\fontsize{7}{7}\textbf{Dice} \\
        \hline
  
Concatenation &  0.6088&0.7400 &  0.8392& 0.9119 &  0.8047&0.8906\\
Adding & \textbf{0.6489}& \textbf{0.7785}&\textbf{0.8514} & \textbf{0.9183} &\textbf{0.8069} &\textbf{0.8917}\\
        \hline
    \end{tabular}
\end{table}
    \begin{table*}[ht]
    \caption{Experimental Results of Different Nonlinearity Decreasing Methods for LV-UNet (Deployment Mode) }
    \label{comp}
    \centering
    \setlength{\tabcolsep}{3.0mm}
\begin{tabular}{c|c|c|c|c|c|c|c|c|c|c}
    \hline
 \multirow{2}{*}{\textbf{Methods}}&\multicolumn{2}{|c|}{\textbf{ ISIC 2016 }} & \multicolumn{2}{|c|}{\textbf{ BUSI}} &  \multicolumn{2}{|c|} {\textbf{ CVC-ClinicDB}} &  \multicolumn{2}{|c|} {\textbf{CVC-ColonDB}} &  \multicolumn{2}{|c} {\textbf{ Kvasir-SEG}} \\
   \cline{2-11} &\textbf{IoU} &\textbf{Dice} & \textbf{IoU} &\textbf{Dice}& \textbf{IoU} &\textbf{Dice} &   \textbf{IoU} &\textbf{Dice}&   \textbf{IoU} &\textbf{Dice}\\
        \hline
  
$a={e}/{E}$ & 0.8575& 0.9223& 0.6293 & 0.7609 &\textbf{0.8300} & \textbf{0.9053}&\textbf{0.7983}&0.8810&0.7952&0.8834\\
$a=1-cos({\pi e}/{2E})
$  & \textbf{0.8593}&  \textbf{0.9234}& \textbf{0.6347} &  \textbf{0.7642} &0.8272 &0.9045&0.7975&\textbf{0.8811}&\textbf{0.7956}&\textbf{0.8840}\\

        \hline
    \end{tabular}
\end{table*}

\textbf{The Way of Skip-Connections.} We evaluated the two types of skip-connections methods: adding and concatenation. The adding operation in skip-connections outperforms concatenation on all three datasets while reducing parameters and computational complexities. The comparison results can be seen in Table \Rmnum {6}.

\textbf{Deep Training Strategies.}
We also compared different nonlinearity decreasing methods: $a=1-cos({\pi e}/{2E})
$ and the design in ~\cite{chen2023vanillanet} $a={e}/{E}$. The comparison results in Table \Rmnum {7} show that the performances are very close and that our method achieves better results in more datasets.

\section{Conclusions}
This paper proposes a lightweight and vanilla model named LV-UNet, which utilizes a pre-trained MobileNetv3-Large model and introduces fusible modules. The model can be trained using an improved deep training strategy and shifted to deployment mode during inference, reducing both the parameter count and computational load. Experiments are conducted on ISIC 2016, BUSI, CVC-ClinicDB, CVC-ColonDB and Kvair-SEG datasets, achieving better performance compared to state-of-the-art and classic models. The effect of the number of pre-trained layers, the number of series-informed activation functions, the way of skip-connections and deep training strategies are carefully studied. The research presents a general way to design a lightweight model for medical image segmentation: the combination of pre-trained models and fusible modules, which may provide some hints for further studies. 

\textbf{Limitations} The learning rate in our setting might not be optimal for larger models, and experiments were not repeated multiple times. Besides, utilizing a more advanced pre-trained model could be a promising direction, as MobileNetv3-Large is no longer the state-of-the-art lightweight model.

\section*{Acknowledgment}
This work was supported by a Grant from The National Natural Science Foundation of China (No. 62103363).

\bibliographystyle{IEEEtran}
\bibliography{IEEEexample}

\end{document}